\documentclass[10pt]{article}
\usepackage[latin9]{inputenc}
\usepackage{color}
\usepackage{amsmath}
\usepackage{amssymb}
\usepackage{graphicx}

\makeatletter
\@ifundefined{date}{}{\date{}}
\usepackage{cite}

\usepackage[labelfont=bf,labelsep=period,justification=raggedright]{caption}

\renewcommand{\@biblabel}[1]{\quad#1.}

\makeatother

\begin{document}

\begin{flushleft}
\textbf{\Large{}Toward Biochemical Probabilistic Computation}{\Large{} } 
\\
Jacques Droulez$^{a,b,\ast}$, 
David Colliaux$^{a}$, 
Audrey Houillon$^{b,}\footnote{At present, Bernstein Center for Computational Neuroscience, Berlin, Germany}$,
Pierre Bessière$^{a,b}$
\\
\textbf{{a} }CNRS/Sorbonne Universités/UPMC/ISIR, Paris, France
\\
\textbf{{b} }CNRS/Collège de France/LPPA, Paris, France
\\
$\ast$ E-mail: Jacques.Droulez@College-de-France.fr 
\par\end{flushleft}


\section{\textmd{Abstract}}

To account for the ability of living organisms to reason with uncertain
and incomplete information, it has been recently proposed that the
brain is a probabilistic inference machine, evaluating subjective
probabilistic models over cognitively relevant variables. A number
of such Bayesian models have been shown to account efficiently for
perceptive and behavioral tasks. However, little is known about the
way these subjective probabilities are represented and processed in
the brain. Several theoretical proposals have been made, from large
populations of neurons to specialized cortical microcircuits or individual
neurons as potential substrates for such subjective probabilistic inferences.
In contrast, we propose in this paper that at a subcellular level,
biochemical cascades of cell signaling can perform the necessary probabilistic
computations. Specifically, we propose that macromolecular assemblies
(receptors, ionic channels, and allosteric enzymes) coupled through several
diffusible messengers (G-proteins, cytoplasmic calcium, cyclic nucleotides
and other second messengers,  membrane potentials, and neurotransmitters)
are the biochemical substrates for subjective probability evaluation
and updating. On one hand, the messengers' concentrations play the
role of parameters encoding probability distributions; on the other
hand, allosteric conformational changes compute the probabilistic
inferences. The method used to support this thesis is to prove
that both subjective cognitive probabilistic models and the descriptive
coupled Markov chains used to model these biochemical cascades are
performing equivalent computations. On one hand, we demonstrate that
Bayesian inference on subjective models is equivalent to the computation
of some rational function with nonnegative coefficient (RFNC), and,
on the other hand, that biochemical cascades may also be seen as computing
RFNCs. This suggests that the ability to perform probabilistic reasoning
is a very fundamental characteristic of biological systems, from unicellular
organisms to the most complex brains.


\section{\textmd{Author Summary}}

Living organisms survive and multiply even though they have uncertain
and incomplete information about their environment and imperfect models
to predict the consequences of their actions. Bayesian models have
been proposed to face this challenge. Indeed, Bayesian inference is
a way to do optimal reasoning when only uncertain and incomplete
information is available. Various perceptive, sensory-motor, and cognitive
functions have been successfully modeled this way. However, the biological
mechanisms allowing animals and humans to represent and to compute
probability distributions are not known. It has been proposed that
neurons and assemblies of neurons could be the appropriate scale to search
for clues to probabilistic reasoning. In contrast, in this paper,
we propose that interacting populations of macromolecules and diffusible messengers
can perform probabilistic computation. This suggests
that probabilistic reasoning, based on cellular signaling pathways,
is a fundamental skill of living organisms available to
the simplest unicellular organisms as well as the most complex brains.

\section{\textmd{Introduction}}

The information available to living organisms about their environment
is uncertain, not only because biological sensors are imperfect, but
more importantly because sensors inevitably provide an incomplete,
partial description of the environment. Moreover, timing is a crucial
constraint for biological systems. During a fight-or-flight dilemma,
animals must quickly decide to fight or flee, and they can never be sure of the good
or bad consequences of their decision. Incompleteness is therefore
a key notion for an autonomous agent facing the complexity of the
world. There is now growing evidence that probabilistic reasoning
is a rigorous and efficient way to represent partial or incomplete
knowledge and to answer questions optimally that have no uniquely
defined solution \cite{jaynes:03}.

Perception is a well-known example of ill-posed problems, because an
indefinite number of object characteristics can theoretically give
rise to the same set of sensory data. For instance, an indefinite
number of objects with various shapes, sizes and movements can induce
exactly the same retinal projection. A number of studies have shown
that Bayesian models can accurately account for various aspects of
perception \cite{knill:96, Weiss:2002:Motion-ill:g, Rao:2002:Probabilis:p, mamassian:02, Ernst:2002:Humans-int:d, laurens:07:a} and sensory-motor
integration \cite{kording:04, Kording:2006:Bayesian-d:h, COLAS:2008:HAL-00215918:1, Todorov:2008:General-du:q}.

Nonprobabilistic models assume that the brain computes internal estimates
of relevant state variables such as motion, object distance, and color.
Each variable is supposed to have a unique estimate, on which
no evaluation of uncertainty is performed. In contrast, Bayesian models
assume that the brain evaluates the probability corresponding to each
possible value of the relevant variables. Probability computation
results from straightforward application of the Bayes (\textit{i.e.}, multiplicative)
and marginalization (\textit{i.e.}, additive) rules, which can been seen as
a generalization of logical inference to probability distributions
\cite{Bessiere:2008:Probabilis:k, Bessiere:2013:Bayesian-P:r}. Following Pearl \cite{pearl:88} and Jaynes \cite{jaynes:03},
we define probabilistic reasoning as the ability to perform
inference within a probabilistic knowledge base. In the following,
we will call a subjective Bayesian model the specification of the
variables of interest, their conditional dependencies, the parametric
forms, and the way probability distributions can be inferred.

\bigskip{}

\noindent Acknowledging the efficiency of probabilistic reasoning
in accounting for a large variety of perceptive reports or motor behaviors,
one of the main scientific challenges is to explicitly demonstrate
that the brain, and more generally biological systems, can effectively
perform probabilistic computation. The problem is to show possible
correspondences between subjective Bayesian models and descriptive
models of biological systems and signal processing.

\bigskip{}

\noindent Most existing studies proposed assemblies of neuronal
cells or single neurons as the appropriate level of analysis to explain
how a brain could perform probabilistic inference. Several authors have
proposed that the firing rate of a group of cells within a given temporal
window \cite{Zemel:1998:Probabilis:c, Deneve:1999:Reading-po:g, Gold:2002:Banburismu:d} could represent probability
distributions over state variables. The mean firing rate might be
well approximated by a graded value, to which a subjective probabilistic
meaning can be attributed: for instance, the probability that a given
proposal over a state variable is true, or its log-likelihood ratio.
Other approaches are based on the Poisson-like variability of spike
trains \cite{Ma:2008:Spiking-ne:e} or kernel convolution for encoding/decoding
spike trains \cite{Deneve:2008:Bayesian-S:b}.

\bigskip{}

\noindent In contrast, we consider the molecular scale
as an adequate framework to solve this matching problem between subjective
Bayesian models and descriptive biological ones.

Populations of macromolecules in their various conformational states
and diffusible messenger concentrations are assumed to be the substates
used at subcellular level to represent and to compute probability
distributions. Our proposal is that the biochemical processes involved
in cell signaling can perform the elementary computations needed for
subjective probabilistic reasoning, and that this biochemical computation
is used extensively by the brain as an elementary component: a Bayesian
``nanogate.'' Specifically, we propose that macromolecular assemblies
(receptors, ionic channels, and allosteric enzymes) coupled through several
diffusible messengers (G-proteins, cytoplasmic calcium, cyclic nucleotides
and other second messengers, membrane potentials, and neurotransmitters)
are the biochemical substrates for subjective probability evaluation
and updating. On the one hand, messenger concentrations play the
role of parameters encoding probability distributions, on the other
hand, allosteric conformational changes compute the probabilistic
inferences. Diffusible messenger concentrations, including electric
charge density, control the probability of conformational changes,
which are in turn responsible for inflow and outflow rates of messengers
and are then controlling their kinetics.

We started exploring these ideas in \cite{Houillon2010}, where
the kinetics of a rhodopsin channel were shown to converge to a posterior distribution of states in a
hidden Markov model, the hidden state representing the presence or absence of light.
A similar
kinetic scheme derived from Bayesian formulation of receptor activations was compared with other
common signaling schemes in \cite{Kobayashi2010} to emphasize its optimality. Rather than considering a binary
(ON or OFF) hidden state, multiple activated states were considered in \cite{Siggia2013} (like self and nonself
in the immune system) and they discussed how Koshland--Goldbeter could solve this decision
problem. Napp and Adams proposed in a recent paper \cite{NIPS2013_4901} a procedure to compile a probabilistic graphical model into a chemical reaction network. 


\section{\textmd{Results}}

\emph{Subjective probabilistic models} (often called Bayesian models)
include variables describing states of the world highly relevant for
the organism, such as the presence of food or predators. The values
taken by these variables cannot be known with certainty by the organism.
However, according to the Bayesian approach, the organism can evaluate
the probability distribution over these variables on the basis of
specific observations, such as the detection of light or the detection
of an odorant molecule.

Subjective probabilistic reasoning is the process by which the probabilities
of relevant variables are computed from a set of \emph{observations}
and a set of \emph{priors} and \emph{conditional dependencies} between
variables \cite{Bessiere:2013:Bayesian-P:r}.

\bigskip{}

\noindent \emph{Descriptive biochemical models}, reduce biochemical
signaling mechanisms to two sets of interacting substrates: (i)
\emph{macromolecular assemblies} (such as receptors, ionic channels,
or activatable enzymes), which have a nearly fixed spatial distribution
(in the time scale we considered here), but a variable conformational
state distribution; and (ii) \emph{diffusible messengers} (such as
G-protein $\alpha$-subunits, cyclic nucleotides, cytoplasmic calcium,
and membrane potentials) that have a fixed conformational state, but
a variable spatial distribution.

The messengers, when they bind to their specific receptor sites on
macromolecules, induce conformational changes. The macromolecules,
when they are in active states, release or remove messengers, thus
changing the messengers' spatial distribution.  Continuous thermal agitation
induces random events such as macromolecule allosteric transitions,
messenger diffusion, and molecular collisions. Therefore, biochemical
interactions can be described as a set of strongly coupled causal
Markov chains, with random variables specifying internal states such as
conformational states and concentrations.

\bigskip{}

\noindent The main result of this paper is a formal demonstration
that concentrations of messengers in biochemical systems can represent
subjective probability distributions and that conformational changes
of macromolecules can perform the fundamental operations required
by Bayesian inference.

To reach this goal we demonstrate, on one hand, that Bayesian inference
on subjective models is equivalent to the computation of some rational
function with nonnegative coefficient (RFNC) of inputs related to
sensory data, and, on the other hand, that biochemical cascades may
also be seen as computing RFNCs of inputs related to the concentration
of primary messengers.

\subsection{\textmd{A basic example}}

The simplest subjective model has only one binary variable $S\in\{0,\,1\}$
and a single observation $K\in\mathbb{K}$.
The structure of the model consists of a prior on $S$, $P(S)$, and
a likelihood of the observation $K$ knowing $S$, $P(K|S)$: 
\begin{equation}
P(S,K)=P(S)P(K|S)
\end{equation}

The prior on $S$ is completely specified by the odds: 
\begin{equation}
\frac{P([S=1])}{P([S=0])}=a
\end{equation}

It is straightforward to see that $P([S=0])=1/(1+a)$ and $P([S=1])=a/(1+a)$.

The likelihood function $P(K\,|\, S)$ allows us to define a mapping
$g$ from $\mathbb{K}$ to $\mathbb{R}^{+}$such that:

\begin{equation}
g(k)=\frac{P([K=k]\,|\,[S=1])}{P([K=k]\,|\,[S=0])}
\end{equation}

For a given observation $k$, we apply this function to find the input
number $x=g(k)$, from which we can compute the corresponding posterior
odds:

\begin{equation}
y=\frac{P([S=1]\,|\, k)}{P([S=0]\,|\, k)}=ax
\end{equation}

Obviously, this single nonnegative number $y$ completely specifies
the posterior distribution with $P([S=0]\,|\, k)=1/(1+y)$ and $P([S=1]\,|\, k)=y/(1+y)$.

This extremely basic subjective model illustrates the two main operations
required by probabilistic inference: once an observation $k$ is known,
it is first transformed into a nonnegative number $x$ that we call
the input (we will later show that in the general case the input is
a multidimensional vector of nonnegative numbers). The input
is then processed by an RFNC,
reduced in this basic model to a multiplication by a constant. The
output is again a nonnegative number (in the general case, a multidimensional
vector of nonnegative numbers), which specifies the posterior distribution.

\bigskip{}

\begin{figure}[htb]
\includegraphics[width=.3\linewidth]{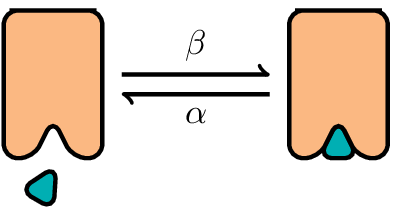}
\includegraphics[width=.65\linewidth]{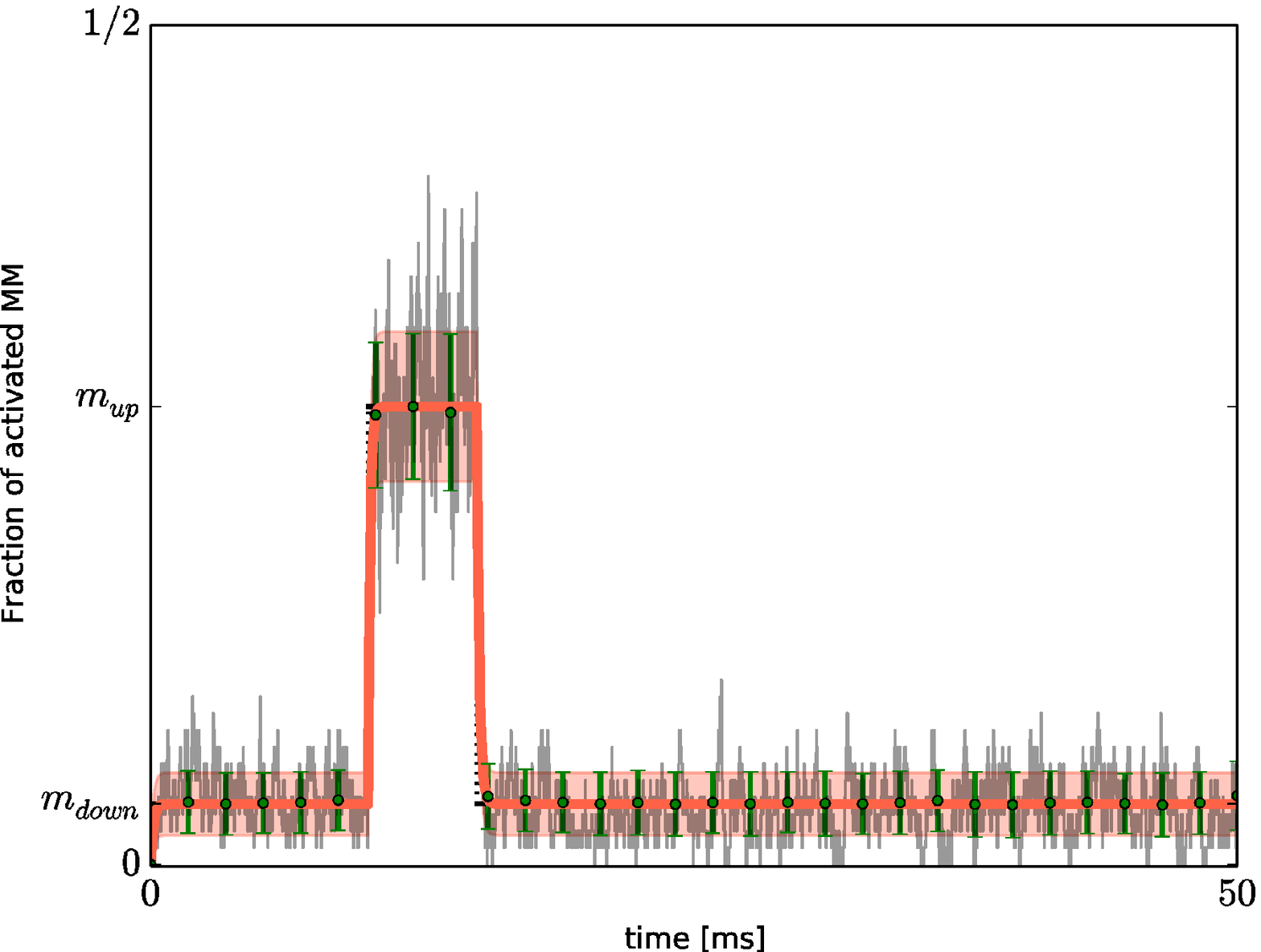}
\caption{(Left) Schema of the reaction involved in a single receptor site. The two constants $\alpha, \beta$ are the kinetic constants for the release and fixation of the messenger, respectively. (Right) Dynamics of receptor binding when the ligand concentration is increased for 5 ms for a population of
100 receptors. The gray line shows the fraction of bound molecules as simulated with the Gillespie
algorithm. The green dots show the ensemble average ($\pm$  one standard deviation) of these
trajectories over 500 repetitions. The orange curve shows the same
variable as obtained by the continuous time dynamics of the mean and
variance (see Section \ref{sec:parameters}  for the derivation of these equations). Parameters are chosen according to the acetylcholine
receptor described in \cite{Edelstein:1996:A-kinetic-:f} (see Table \ref{param_table}).}
\label{fig:one-receptor-site} 
\end{figure}

\noindent We will now show how these two successive operations can
be implemented by a biochemical system (see Figure \ref{fig:one-receptor-site}).
We first consider a population of macromolecular receptors sensitive
to some external stimulus. Depending on the type of receptor, the
stimulus could be, for instance, a flow of photons, a mechanical strength,
or the presence of nutrient in the environment. This stimulus is
what the cell \emph{observes} from its current environment, and its
actual particular value can be denoted as $k$. This stimulus induces
a conformational change, and thus the proportion of active receptors.
The activity of these receptors results in the release of a primary
messenger $X$ within the cell. We call $x=[X]$ the concentration
of the primary messenger resulting from the stimulus $k$. The population
of receptors plays exactly the role of the mapping $g$ in the subjective
model. Although $k$ is not represented \textit{per se} in the biochemical
system, the output of the population of receptors, \textit{i.e.}, the
primary messenger concentration $x$, represents what really matters
for the inference, the likelihood ratio $g(k)$.

We next consider a second population of macromolecules $M$, each with a
single receptor site, specific for the primary messenger $X$. These
macromolecules could be in one of two conformations: in the first, the
receptor site is free ($M=0$) and in the other the receptor binds a
messenger ($M=1$). Within a small time interval $\Delta t$,  the
transition probability from $M=1$ to $M=0$ is a constant $\alpha$.
The transition from $M=0$ to $M=1$ requires the presence of one
messenger molecule in the vicinity of the receptor site; its probability
is therefore proportional to the current messenger concentration $x$:
\begin{equation}
\begin{cases}
\,\, P_{1\rightarrow0}=P(M^{t+\Delta t}=0\,|\, M^{t}=1,\, x)=\alpha\\
\,\, P_{0\rightarrow1}=P(M^{t+\Delta t}=1\,|\, M^{t}=0,\, x)=\beta x
\end{cases}
\end{equation}
Each macromolecule in the $M$ population switches randomly between
state 0 and state 1 according to the transition probabilities defined
above. After a while, the probability of finding the macromolecule in
a given state converges to the equilibrium, or steady-state, distribution.
In this case, the equilibrium distribution ($P^{*}$) follows the detailed
balance equation: 
\begin{equation}
P^{*}_{1}.P_{1\rightarrow0}=P^{*}_{0}.P_{0\rightarrow1}
\end{equation}
and the resulting equilibrium distribution is: 
\begin{equation}
\begin{cases}
\,\, P^{*}_{0}=\alpha/(\alpha+\beta x)\\
\,\, P^{*}_{1}=\beta x/(\alpha+\beta x)
\end{cases}
\end{equation}
When in state 1, that is, when the primary messenger is fixed, the
macromolecules release a second messenger $Y$ (concentration $y=[Y]$)
at a constant rate $a_{P}$. When in state 0, that is, when the receptor
site is free, the macromolecules remove the second messenger at a
rate proportional to its concentration $a_{R}\times y$, so that
at equilibrium we have: 
\begin{equation}
a_{P}\times\frac{\beta x}{\alpha+\beta x}=a_{R}\times\frac{\alpha}{\alpha+\beta x}\times y
\end{equation}
Finally, the resulting second messenger concentration is simply proportional
to the primary messenger concentration: 
\begin{equation}
y=\frac{a_{P}}{a_{R}}\times\frac{\beta x}{\alpha}
\end{equation}

\bigskip{}

\noindent This basic example illustrates the kind of equivalence scheme
that can be drawn between the subjective and descriptive
models. The macromolecular receptors transform a stimulus $k$ into
a likelihood ratio represented by the concentration of primary messengers,
while the remainder of the biochemical chain computes a posterior
ratio represented by the concentration of second messengers. Obviously,
this basic model is quite elementary (see a simulation with realistic
values in Figure \ref{fig:one-receptor-site}). In the next two sections,
we will show that the equivalence scheme is much more general.

\subsection{\textmd{Bayesian models and RFNCs}}

In this paper, we consider only subjective models where unknown variables
are discrete,\emph{ i.e.}, they have a finite, though eventually huge,
number of possible values. The unknown variables can be divided into
a set of relevant, or searched variables, denoted $S$ (of size $n_{S}$)
and a set, eventually empty, of intermediate, or free variables, denoted
$F$ (of size $n_{F}$). We do not make any restriction on the nature,
discrete or not, of the set of observations $K$. Based on these assumptions,
the structure of the general subjective model we consider is: 
\begin{equation}
P(S,\, F,\, K)=P(S,\, F)\times P(K\,|\, S,\, F)
\end{equation}
The right side of this equation is the product of a prior distribution
$P(S,\, F)$ and a likelihood $P(K\,|\, S,\, F)$. For the organism,
the problem is to compute the posterior distribution over the relevant
variable $S$ knowing a given set of observations $k$ (by convention,
we will use small caps for variables with known values): 
\begin{equation}
P(S\,|\, k)=\frac{\sum_{F}\, P(S,\, F)\times P(k\,|\, S,\, F)}{\sum_{S}\sum_{F}\, P(S,\, F)\times P(k\,|\, S,\, F)}
\end{equation}
The denominator on the right side of the equation is a normalization
constant, which could be hard to compute if the state
spaces ($S$ and $F)$ are very large. Instead, the posterior distribution
is completely determined by a finite set of probability ratios: 
\begin{equation}
\frac{P([S=s]\,|\, k)}{P([S=0]\,|\, k)}=\frac{\sum_{F}\, P([S=s],\, F)\times P(k\,|\,[S=s],\, F)}{\sum_{F}\, P([S=0],\, F)\times P(k\,|\,[S=0],\, F)}
\end{equation}
where the particular state $[S=0]$ is used as a reference, or default
state. One can also choose a reference value for the free variables.
Providing that the prior $P([S=0],\,[F=0])$ and the likelihood $P(k\,|\,[S=0],\,[F=0])$
for the chosen reference values are not null, the posterior ratios
reduce to: 
\begin{equation}
\frac{P([S=s]\,|\, k)}{P([S=0]\,|\, k)}=\frac{\sum_{f}\,\frac{P([S=s],\,[F=f])}{P([S=0]\,[F=0])}.\frac{P(k\,|\,[S=s],\,[F=f])}{P(k\,|\,[S=0]\,[F=0])}}{\sum_{f}\,\frac{P([S=0],\,[F=f])}{P([S=0],\,[F=0])}.\frac{P(k\,|\,[S=0],\,[F=f])}{P(k\,|\,[S=0],\,[F=0])}}\label{eq:posterior-ratio}
\end{equation}

\subsubsection{\textmd{From Bayesian models to RFNCs}}

\noindent Once a given observation $k$ is available, it can be first
transformed into an input vector $x=g(k)$ of dimension $n_{SF} = n_{S} n_{F}$ 
with the following mapping: 
\begin{equation}
\forall s\in\{0,...,\, n_{S}-1),\,\forall f\in\{0,...,\, n_{F}-1):\, x_{s,f}=\frac{P(k\,|\,[S=s]\,[F=f])}{P(k\,|\,[S=0]\,[F=0])}
\end{equation}

Obviously, we have $x_{0,0}=1$ and all other components are nonnegative
numbers.

Similarly, the prior $P(S,\, F)$ is completely specified by a vector
of nonnegative constants $a$ such that: 
\begin{equation}
\forall s\in\{0,...,\, n_{S}-1),\,\forall f\in\{0,...,\, n_{F}-1):\, a_{s,f}=\frac{P([S=s]\,[F=f])}{P([S=0]\,[F=0])}
\end{equation}

Finally, the posterior $P(S\,|\, k)$ is completely specified by a
vector of dimension $n_{S}$: 
\begin{equation}
\forall s\in\{0,...,\, n_{S}-1):\, y_{s}=\frac{P([S=s]\,|\, k)}{P([S=0]\,|\, k)}
\end{equation}

Equation (\ref{eq:posterior-ratio}) can be restated as: 
\begin{equation}
\forall s:\, y_{s}=\frac{\sum_{f}\, a_{s,f}\times x_{s,f}}{\sum_{f}\, a_{0,f}\times x_{0,f}}
\end{equation}

Obviously, we have $y_{0}=1$ and all other components are rational
functions with nonnegative coefficients of the components of the
input vector.

\bigskip{}

This demonstrates that the exact inference in the Bayesian model is equivalent
to the application of a finite set of 
RFNCs to an observation-dependent vector of nonnegative
numbers.

\subsubsection{\textmd{From RFNC to Bayesian models}}

\label{From RFNC to Bayesian models}

Reciprocally, we will now show that from any finite set of RFNCs, one
can build at least one subjective probabilistic model such that the
posterior probability ratios can be computed with these RFNCs.

\bigskip{}

\noindent Let us first prove this result for a single RFNC $h$. Clearly,
in this case, the search variable $S$ is a binary random variable,
and the problem is to specify the subjective model such that the posterior
odds is equal to $h(x)$ where $x$ is an input vector depending on
the set of observations or known variables $k$.

We define the complexity of an RFNC as the number of operations (addition,
multiplication, and division) required to compute the image $h(x)$
from the components of the input vector.

If $complexity(h)=0$, $h(x)$ is either a constant or one component
of the input vector. There are obvious Bayesian models associated
with both cases, as shown in the basic model section.

If $complexity(h) \geq 1$, then it is possible to decompose $h$ into
a combination of two simpler, (\textit{i.e.}, lower-complexity) RFNCs
$h_{1}$ and $h_{2}$ with either $h(x)=h_{1}(x)+h_{2}(x)$, $h(x)=h_{1}(x).h_{2}(x)$,
or $h(x)=h_{1}(x)/h_{2}(x)$.

Suppose that we can associate $h_{1}$ with a probabilistic model $P(S_{1},\, F_{1},\, K)$,
where the searched variable $S_{1}$ is binary, such that: 
\begin{equation}
\frac{P([S_{1}=1]\,|\, k)}{P([S_{1}=0]\,|\, k)}=h_{1}(x)
\end{equation}

Similarly, we suppose that we can associate $h_{2}$ with another probabilistic
model $P(S_{2},\, F_{2},\, K)$ where $S_{2}$ is binary such that
the posterior ratio is equal to $h_{2}(x)$.

We may assemble these two models using a generative metamodel defined
by:

\begin{equation}
\begin{array}{ll}
 & P(S,\, S_{1},\, S_{2},\, F_{1},\, F_{2},\, K,\, K_{1},\, K_{2},\,\Lambda,\,\Gamma)\\
= & P(S)P(S_{1},\, F_{1},\, K_{1})P(S_{2},\, F_{2},\, K_{2})P(\Lambda\,|\, S,\, S_{1},\, S_{2})P(\Gamma\,|\, K,\, K_{1},\, K_{2})
\end{array}
\end{equation}

This metamodel uses two binary variables $\Lambda$ and $\Gamma$, which are
called coherence variables (see Chapter 8 in \cite{Bessiere:2013:Bayesian-P:r}) and define how the two submodels
are combined.

The distribution $P(\Gamma\,|\, K,\, K_{1},\, K_{2})$ is defined
as a Dirac distribution, which means that the same set of observations
$K$ is shared by both submodels: 
\begin{equation}
P([\Gamma=1]\,|\, K,\, K_{1},\, K_{2})=1_{K=K_{1}=K_{2}}
\end{equation}

The distribution $P(\Lambda\,|\, S,\, S_{1},\, S_{2})$ is defined
as follows: 
\begin{equation}
P([\Lambda=1]\,|\,[S=i],\,[S_{1}=j],\,[S_{2}=k])=q_{i,j,k}
\end{equation}
where the coefficients $q_{i,j,k}$ define how $S$ is related to
$S_{1}$ and $S_{2}$.

We may now consider the model specified by: 
\begin{equation}
P(S,\, S_{1},\, S_{2},\, F_{1},\, F_{2},\, K,\, K_{1},\, K_{2}|\left[\Lambda=1\right],\left[\Gamma=1\right])
\end{equation}

The posterior ratio on $S$ knowing $k$ for this model is equal to:
\begin{equation}
\frac{P([S=1]\,|\, k,\,[\Lambda=1],\,[\Gamma=1])}{P([S=0]\,|\, k,\,[\Lambda=1],\,[\Gamma=1])}=\frac{q_{1,0,0}+q_{1,1,0}h_{1}(x)+q_{1,0,1}h_{2}(x)+q_{1,1,1}h_{1}(x)h_{2}(x)}{q_{0,0,0}+q_{0,1,0}h_{1}(x)+q_{0,0,1}h_{2}(x)+q_{0,1,1}h_{1}(x)h_{2}(x)}
\end{equation}
and we can choose the coefficients $q_{i,j,k}$ such that the posterior
ratio on $S$ is equal to:
\begin{enumerate}
\item either the sum $h_{1}(x)+h_{2}(x)$ with all coefficients null except
$q_{1,1,0}=q_{1,0,1}=q_{0,0,0}=1$;
\item or the product $h_{1}(x).h_{2}(x)$ with all coefficients null except
$q_{1,1,1}=q_{0,0,0}=1$;
\item or the quotient $h_{1}(x)/h_{2}(x)$ with all coefficients null except
$q_{1,1,0}=q_{0,0,1}=1$.
\end{enumerate}
Finally, using this generative procedure recursively to derive submodels
of null complexity, we can construct for any $h$ a Bayesian model
such that: 
\begin{equation}
\frac{P\left(\left[S=1\right]|k\right)}{P\left(\left[S=0\right]|k\right)}=h\left(x\right)
\end{equation}

\bigskip{}

\noindent Consider now a set of $n$ RFNCs. For each RFNC $h_{i}(x)$,
we have shown that there exists at least one (generally several) probabilistic
model based on a single searched binary variable $S_{i}$ such that
the posterior ratio is equal to $h_{i}(x)$. We now include
a global discrete variable $S$ that can take $n_{s}=n$ values,
which is related to the binary variables $S_{i}$ by: $S=i\Leftrightarrow S_{i}=1\, and\, S_{j\neq i}=0$.
Again, among other possible encoding schemes (possibly more compact),
this could be specified by a coherence binary variable $\Psi$ such
that $P([\Psi=1]\,|\, S,\, S_{1},...,S_{n})=1$ if and only if when
$S=i$, $S_{j\neq i}=0$ and $S_{i}=1$.

\subsection{\textmd{Biochemical cascades and RFNCs}}
According to the Monod--Wyman--Changeux (MWC) model \cite{Monod196588, Changeux03062005},
the activity of a macromolecule depends on its tridimensional
tertiary or quaternary structure, which can be in a discrete number
of states, typically two, named ``tensed'' and ``relaxed'' in
the original formulation of the model. The transition probability
between these allosteric conformations depends on the status (free
or not) of the receptor sites, and the affinity for specific messengers
depends on the allosteric conformation.

Therefore the state of a macromolecule is defined by the couple $\left(Q,\, R\right)$
where $Q=(Q_{1},...,Q_{n_{Q}})$ is a set of $n_{Q}$ binary variables
specifying the allosteric conformation ($n_{Q}=1$ in the example
of Figure \ref{fig:two-receptor-site}), and $R=(R_{1},...,R_{n_{R}})$
is a set of $n_{R}$ binary variables specifying the state of the
receptor sites ($n_{R}=2$ in the example of Figure \ref{fig:two-receptor-site}).
Each receptor site can bind a specific messenger, so there are
also $n_{R}$ potentially different messenger concentrations represented
by the vector $x=(x_{1},...,x_{n_{R}})$. The macromolecule can take
$2^{n_{M}}$ different states ($n_{M}=n_{Q}+n_{R}$) and for each
of these states, at equilibrium, the probability of leaving the state
is equal to the probability of reaching it: 
\begin{equation}
\forall i:1\leq i\leq2^{n_{M}},P^{*}_{i}\times\sum_{j\neq i}\left[P_{i\rightarrow j}\right]=\sum_{j\neq i}\left[P^{*}_{j}\times P_{j\rightarrow i}\right]
\end{equation}
where $P_{i}$ is the probability of being in state $i$ at equilibrium
and $P_{i\rightarrow j}$ is the probability of switching from state
$i$ to state $j$. Some of the $P_{i\rightarrow j}$ are constants,
while some others assume the presence of a messenger and are consequently
proportional to the concentration $x_{i}$ of this messenger.

Alternatively, we can specify the descriptive model as a Markov chain
with a transition matrix $T$: 
\begin{equation}
P^{t}=T\times P^{t-1}=\left(\begin{array}{ccccc}
P_{1\rightarrow1} &  &  &  & P_{2^{n_{M}}\rightarrow1}\\
 &  &  & P_{i\rightarrow j}\\
\\
P_{1\rightarrow2^{n_{M}}} &  &  &  & P_{2^{n_{M}}\rightarrow2^{n_{M}}}
\end{array}\right)P^{t-1}
\end{equation}
where equilibrium is defined by the eventually existing fixed points:
\begin{equation}
P^{*}=T\times P^{*}
\end{equation}
It should be noted that (i) the coefficient of $T$ are positive
or null and (ii) the coefficients in a column sum
to one because they represent the transition probabilities from one state
to any other state.

\subsubsection{\textmd{From biochemical cascades to RFNC}}

The goal of this section is to demonstrate that biochemical cascades
\emph{without feedback} may be seen as computing RFNCs. Specifically,
we will prove that, when at equilibrium, the concentrations of second
messengers are RFNCs of the concentration of primary messengers.

To reach that goal we will first show that the probability that a macromolecule
in its active state is an RFNC of the concentrations ($x$)
of its primary messengers, and we will then demonstrate that when
a second messenger is both produced and removed by two populations
of macromolecules (see Figure \ref{fig:product}), its concentration ($y$) is
an RFNC of the concentrations of the primary messengers.

\paragraph{Probability for a macromolecule to be in its active state
as an RFNC of $x$:}

In Section \ref{Demo} we presented a general demonstration that the
stationary distributions of a Markov model are RFNCs of the transition
matrix coefficients. As this demonstration is quite technical, the details have
been postponed to the Materials and Methods section. In fact, the coefficients
of transition matrix $T$ of the descriptive model are either constants
or proportional to the primary messenger concentrations. Therefore,
the stationary distribution as well as the probability of being in the
active states are RFNCs of these concentrations.

\begin{figure}[htb]
\includegraphics[width=.3\linewidth]{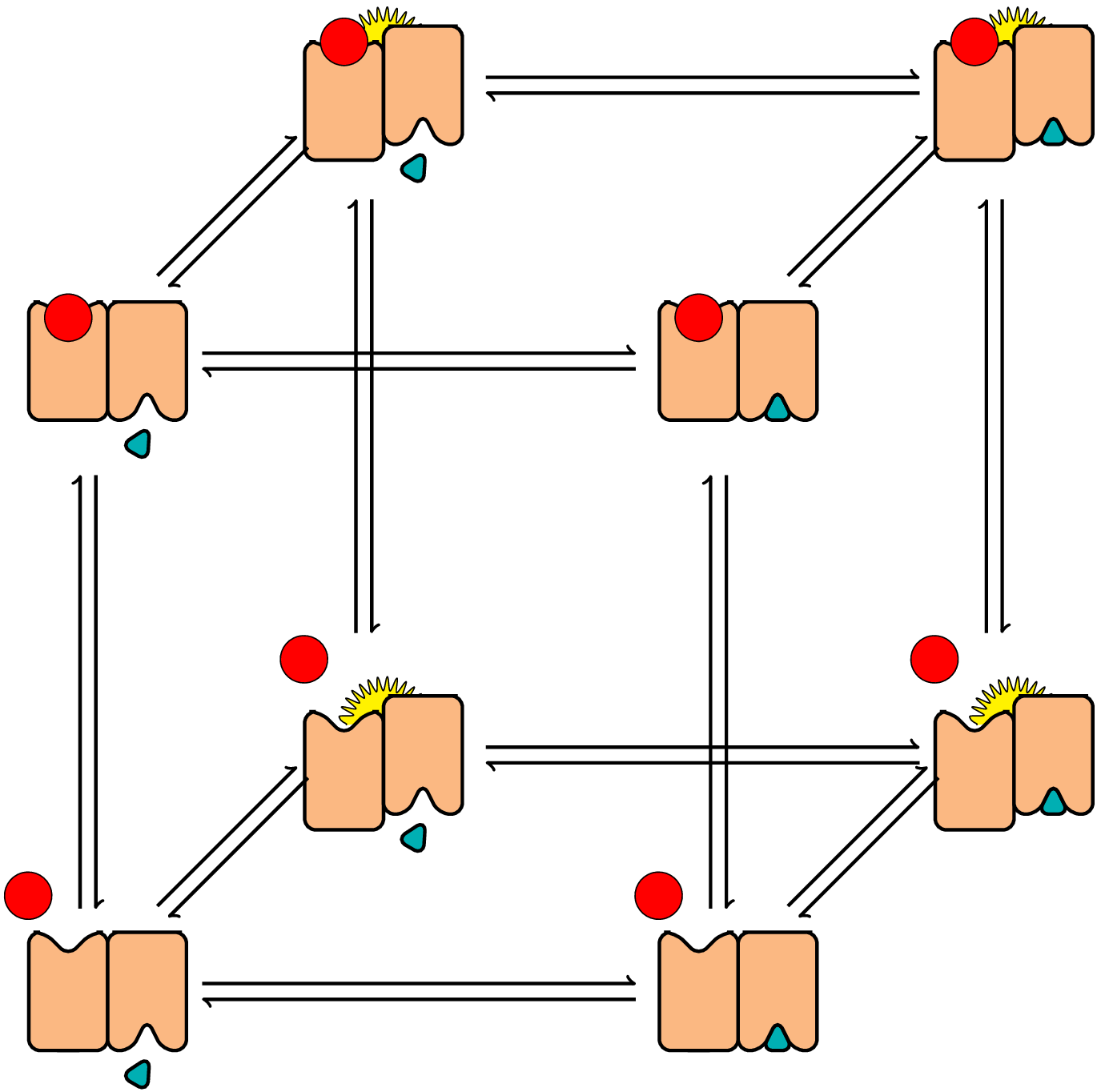}
\includegraphics[width=.65\linewidth]{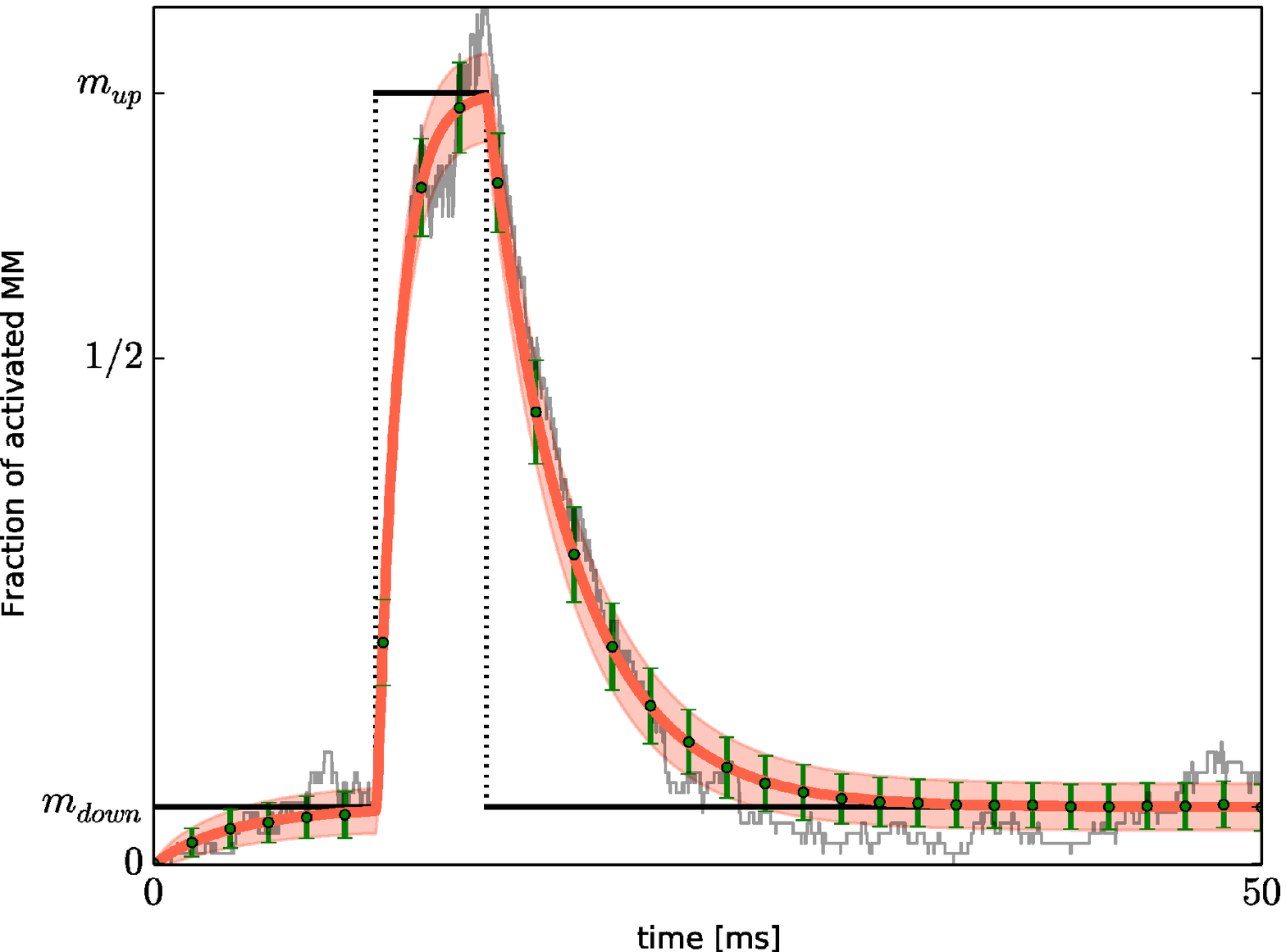}
\caption{(Left) Schema of a macromolecule with two receptor sites (red circle and green triangle) and one catalytic site (yellow). (Right) Simulation of this macromolecule when the concentrations of both ligands are increased from $2\mu M$ to $20\mu M$ during 5 ms. The vertical axis shows the fraction of activated catalytic sites. The same conventions as in Figure \ref{fig:one-receptor-site} are used for the curves. The black line shows the activated fraction of catalytic sites in the stationary state. See Table \ref{param_table_2} for the definition of parameters.}
\label{fig:two-receptor-site} 
\end{figure}

\begin{figure}[htb]
\includegraphics[width=.3\linewidth]{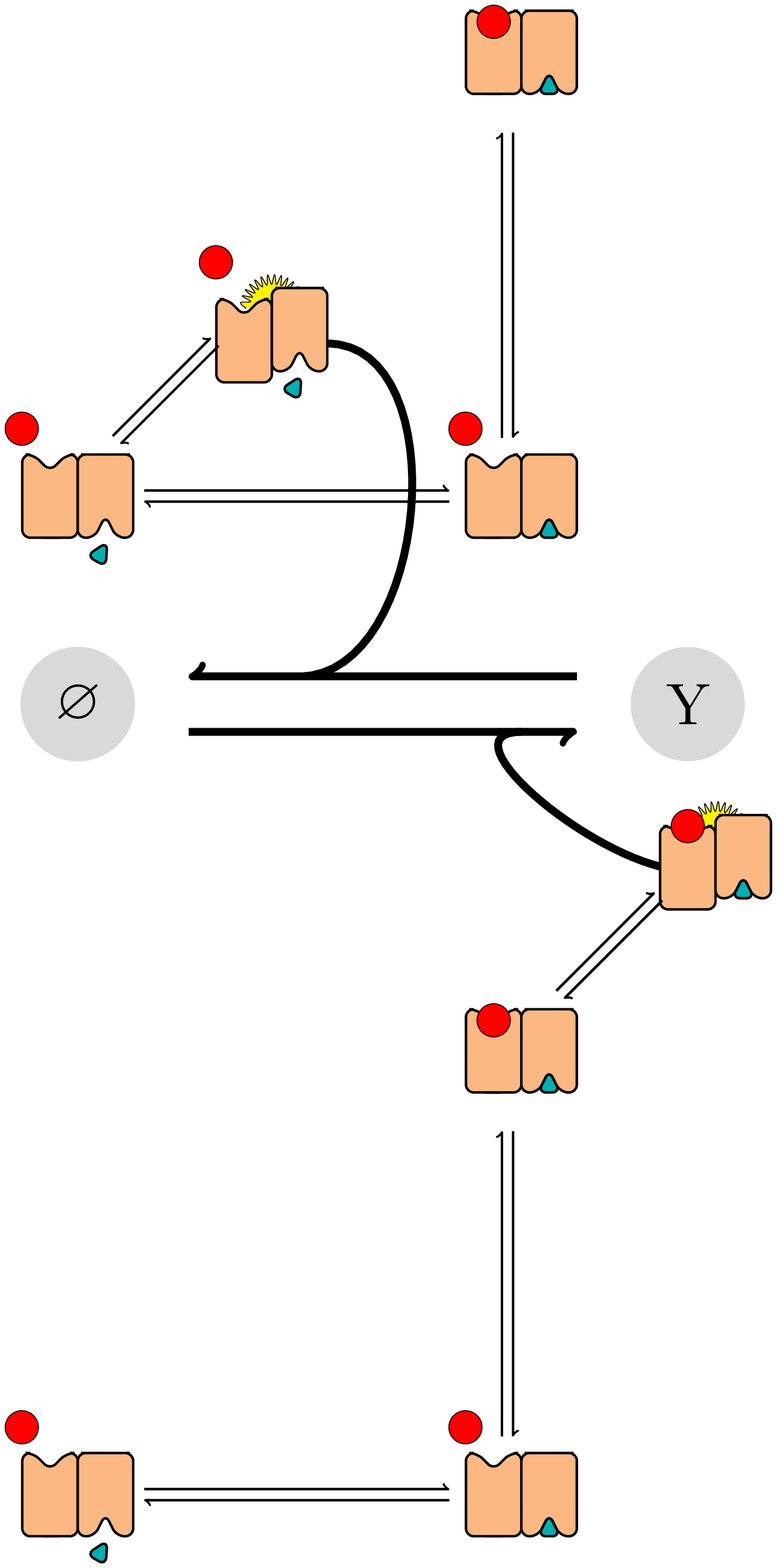}
\includegraphics[width=.65\linewidth]{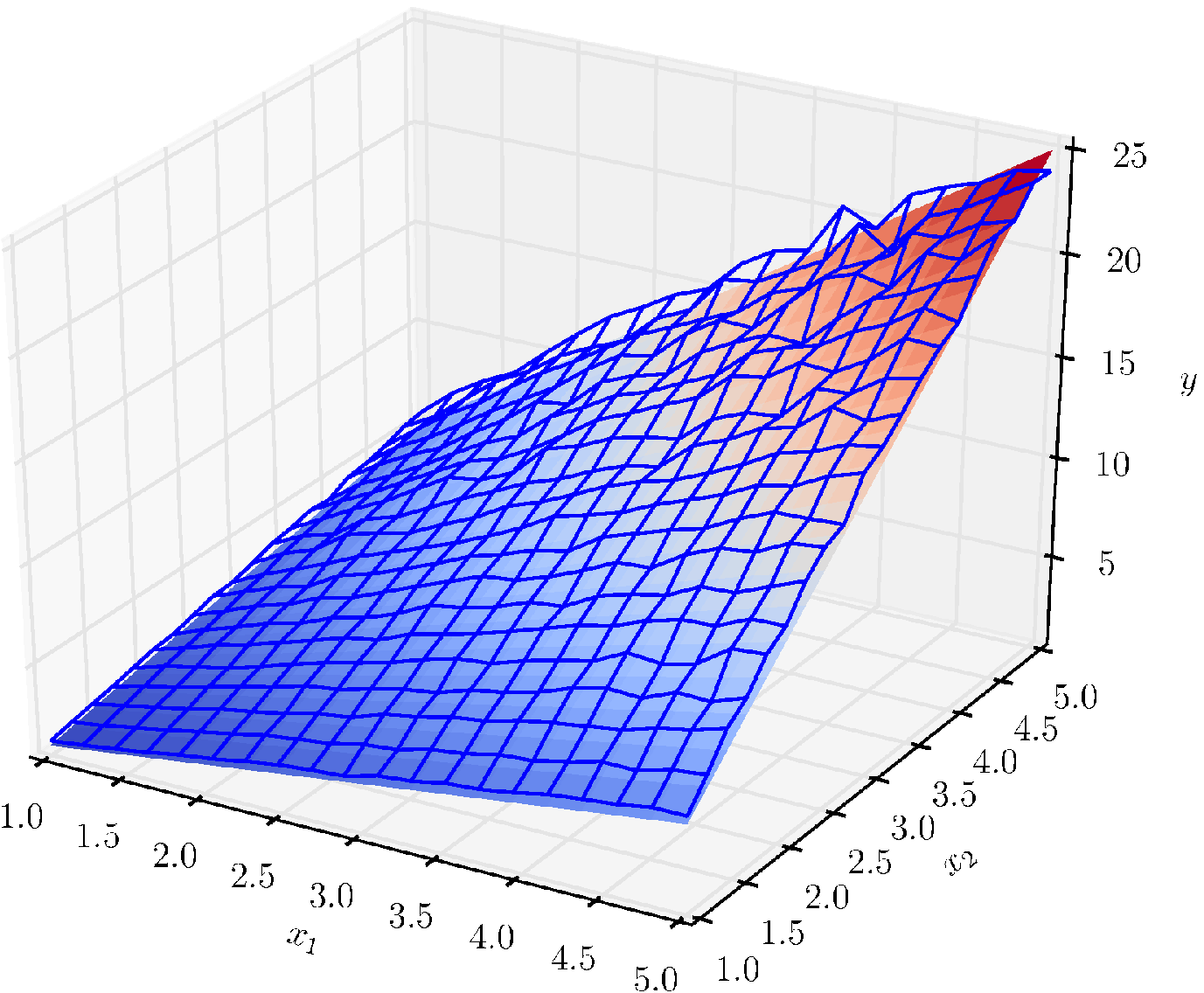}
\caption{(Left) Reaction schema for the production and removal of $Y$ with both reactions controlled by the catalytic site of a macromolecule. The state diagram for each macromolecule is a restricted version of the cube from Figure \ref{fig:two-receptor-site}. Each macromolecule has four  allowed states: three inactive states and one active state. For both macromolecules the three inactive states are : no receptor sites bound, green triangle site bound and both bound. 
The active state of the macromolecule controlling removal of $Y$ (top) has no receptor site bound.
The active state of the macromolecule controlling production of $Y$ (bottom) has both receptor sites bound. 
This cascade implements one of the possible computations listed in Section \ref{sec:RFNCBioc}, the product of the ligands concentrations $y \propto x_1.x_2$. (Right) Result of the simulations where macromolecule transitions are updated using stochastic simulation and $Y$ production is computed using Euler integration. The surface shows the product of ligand concentrations $x_1.x_2$ and
the wireframe shows the scaled simulated concentration in Y.}
\label{fig:product} 
\end{figure}

Here, we only present the demonstration for the special
case of Figure \ref{fig:two-receptor-site} to give a taste of the
complete proof.

The system is made of (i) two primary messengers $X_{1}$ and $X_{2}$
with concentrations $x_{1}$ and $x_{2}$, and (ii) a population of
macromolecules with two receptor sites $R_{1}$ and $R_{2}$ and one
active state $Q$. The allosteric conformation of the macromolecule
is defined by the triplet of three binary values $(R_{1},R_{2},Q)$.

For simplicity, we suppose that transitions in the conformation
space are restrained to switches of a single binary variable at a
time, so that the transition matrix $T$ contains $3\times8=24$ 
nonnull elements (this hypothesis is not necessary for the general
demonstration of Section \ref{Demo}). Each nonnull element of the
transition matrix specifies the probability of changing one particular
binary variable from any given initial conformational state. For the
receptor state variables, the transition from 0 to 1 involves the
presence of one specific messenger in the vicinity of the receptor
site. For the eight transitions of concern here, we have transition probabilities
proportional to the corresponding messenger concentration $x_{i}$;
for instance, $T_{000\rightarrow100}=\alpha_{000\rightarrow100}\times x_{1}$.
The remaining 16 nonnull transition probabilities are constants.

As a second simplifying hypothesis, we suppose that the net chemical
and energetic balance of any complete cycle is null (this hypothesis
is not necessary for the general demonstration of Section \ref{Demo}).
Consequently, the Markov chain is reversible and the conformational
state distribution converges towards an equilibrium distribution $P^{*}(R_{1},R_{2},Q)$, which satisfies the 24 detailed balance equations of which only eight
are independent: 
\begin{equation}
\forall\left(R_{1},R_{2},Q\right)\in\left(0,1\right)^{3}\begin{cases}
P^{*}(R_{1},R_{2},Q)T_{R_{1}R_{2}Q\rightarrow\bar{R_{1}}R_{2}Q}= & P^{*}(\bar{R_{1},}R_{2},Q)T_{\bar{R_{1}}R_{2}Q\rightarrow R_{1}R_{2}Q}\\
P^{*}(R_{1},R_{2},Q)T_{R_{1}R_{2}Q\rightarrow R_{1}\bar{R_{2}}Q}= & P^{*}(R_{1},\bar{R_{2}},Q)T_{R_{1}\bar{R_{2}}Q\rightarrow R_{1}R_{2}Q}\\
P^{*}(R_{1},R_{2},Q)T_{R_{1}R_{2}Q\rightarrow R_{1}R_{2}\bar{Q}}= & P^{*}(R_{1},R_{2},\bar{Q})T_{R_{1}R_{2}\bar{Q}\rightarrow R_{1}R_{2}Q}
\end{cases}
\end{equation}
To compute the probability of a given state at equilibrium we can
follow any path starting from the reference state $(0,0,0)$. For
instance, following the path in Figure \ref{fig:two-receptor-site}: $(0,0,0)$ (no messenger bound), $(1,0,0)$ (green triangle bound), $(1,1,0)$ (both messengers bound), and $(1,1,1)$ (both messengers bound and macromolecule active),
we find: 
\begin{equation}
P^{*}(1,1,1|x_{1},x_{2})=\frac{T_{110\rightarrow111}}{T_{111\rightarrow110}}\frac{T_{100\rightarrow110}}{T_{110\rightarrow100}}\frac{T_{000\rightarrow100}}{T_{100\rightarrow000}}P^{*}(0,0,0|x_{1},x_{2})
\end{equation}
As each transition coefficient is either a constant or a constant
multiplied by one of the two concentrations $x_{1}$ or $x_{2}$, the
probability of each state at equilibrium is an RFNC of the two concentrations.

By marginalizing over $R_{1}$ and $R_{2}$, we obtain the probability
distribution at equilibrium over the activity states: 
\begin{equation}
P^{*}([Q=1]|x_{1},x_{2})=\sum_{R_{1}R_{2}}\left[P^{*}(R_{1},R_{2},[Q=1]|x_{1},x_{2})\right]
\end{equation}
which is also an RFNC.

In real biochemical systems, the proportion of active sites fluctuates
around the mean theoretical value because the population size is finite.
An example simulation is shown in Figure \ref{fig:two-receptor-site}.

\paragraph{Concentration $y$ of a second messenger as an RFNC of
$x$:}

The second messenger production is the net result of a reaction (or
a chain of reactions) catalyzed by the first population of macromolecules
having $m_{P}$ catalytic sites, with a chemical equation of the form:
\begin{equation}
A\rightarrow Y+B
\end{equation}
The precursor $A$ is a substrate (or set of substrates) present in
high concentrations either in the environment or produced in large
amount by the basic cellular activity. Generally, $A$ is not involved
in cell signaling \emph{per se}. The concentration of the optional
metabolite $B$ is supposed to have no incidence on the catalysis
kinetics%
\footnote{However, there are several well-known exceptions to this rule. For
instance, the cleavage of phospholipid PIP2 by the enzyme phospholipase
C leads to the release of two messengers: diacylglycerol  and
inositol triphosphate that in turn target different biochemical
systems.%
}. The production rate $\Phi_{P}$ is the product of the number $m_{P}\times P(Q|x)$
of active macromolecules by the messenger production rate per active
catalytic site%
\footnote{We assume that, at the temporal scale considered here, the catalytic
activity per active site is constant%
} $a_{P}$:
\begin{equation}
\Phi_{P}(x)=a_{P}\times m_{P}\times P^{*}(Q|x)
\end{equation}
which is an RFNC of $x$.

For $y$, the concentration of the second messenger $Y$, to reach
an equilibrium, the cascade must include a removal reaction (see Figure
\ref{fig:product}). The messenger removal is the net result of a reaction (or
a chain of reactions), catalyzed by a second population of macromolecules,
with a chemical equation of the form: 
\begin{equation}
C+Y\rightarrow D
\end{equation}
For $C$ and $D$, the same assumptions can be made
as for $A$ and $B$. However, the fixation of the messenger on the catalytic site
of the removal population requires the presence of messenger in the
compartment, so that the removal rate $\Phi_{R}$ is proportional
to the second messenger concentration: 
\begin{equation}
\Phi_{R}(x)=a_{R}\times m_{R}\times P^{*}(Q'|x)\times y
\end{equation}
where $m_{R}$ is the number of catalytic sites of the second population,
$a_{R}$ is the removal rate, $Q'$ is the active state of the second population
and $x$ is the concentration vector of the primary messengers. $P(Q'|x)$
is an RFNC of $x$.

At equilibrium, the production and removal rate are equal and
we have: 
\begin{equation}
y=\frac{a_{P}\times m_{P}\times P^{*}(Q|x)}{a_{R}\times m_{R}\times P^{*}(Q'|x)}
\end{equation}
$y$ is an RFNC of the concentrations appearing in $x$.

\subsubsection{\textmd{From RFNC to biochemical cascades}}
\label{sec:RFNCBioc}
Reciprocally, in this section we will finally demonstrate that any
RFNC can be computed by a \emph{theoretical} biochemical cascade.
By ``theoretical'' we mean that the kinetic parameters and overall
organization of the cascade necessary to perform the computation of
a given RFNC have absolutely no warranty of biological existence
or even plausibility.

As we already saw in Section \ref{From RFNC to Bayesian models},
any RFNC $h(x)$ can be decomposed into either the sum, product or
quotient of two simpler RFNCs $h_{1}(x)$ and $h_{2}(x)$. Suppose
that for each function, there exists a (theoretical) biochemical cascade
resulting in the release of two second messengers $Y_{1}$ and $Y_{2}$,
and their concentrations at equilibrium are $y_{1}=h_{1}(x)$ and $y_{2}=h_{2}(x)$.
Then, we can define two new macromolecules, both with two receptor
sites specific to $Y_{1}$ and $Y_{2}$; the first macromolecule releasing
and the second removing a new messenger $Z$. The mean catalytic
activity of the first macromolecule is a rational function of $y_{1}$
and $y_{2}$ with the following general parametric form:

\begin{equation}
\Phi_{P}(y_{1},y_{2})=a_{P}\times n\times\frac{a_{0}+a_{1}y_{1}+a_{2}y_{2}+a_{3}y_{1}y_{2}}{(a_{0}+b_{0})+(a_{1}+b_{1})y_{1}+(a_{2}+b_{2})y_{2}+(a_{3}+b_{3})y_{1}y_{2}}
\end{equation}

Similarly, the mean catalytic activity of the removing macromolecule
is a rational function of $y_{1}$ and $y_{2}$ with the following
general parametric form:

\begin{equation}
\Phi_{R}(y_{1},y_{2})=a_{R}\times m\times\frac{a'_{0}+a'_{1}y_{1}+a'_{2}y_{2}+a'_{3}y_{1}y_{2}}{(a'_{0}+b'_{0})+(a'_{1}+b'_{1})y_{1}+(a'_{2}+b'_{2})y_{2}+(a'_{3}+b'_{3})y_{1}y_{2}}
\end{equation}

As we saw above, at equilibrium the concentration of the new messenger
$Z$ is given by:

\begin{equation}
z=\Phi_{P}(y_{1},y_{2})/\Phi_{R}(y_{1},y_{2})
\end{equation}

We will now show that it is possible to find a set of parameters such
that either $z=y_{1}+y_{2}$, $z=y_{1}.y_{2}$, or $z=y_{1}/y_{2}$.
Some, but not all, parameters can be set to zero. These parameters
determine the probability of the macromolecules being in one of the
eight possible states: 
\begin{enumerate}
\item the resting (default) state with no bound messenger and no catalytic
activity is determined by parameters $b_{0}$ and $b'_{0}$; 
\item the inactive state with messenger $Y_{1}$ bound to the receptor site
is determined by parameters $b_{1}$ and $b'_{1}$; 
\item the inactive state with messenger $Y_{2}$ bound to the receptor site
is determined by parameters $b_{2}$ and $b'_{2}$; 
\item the inactive state with both messengers ($Y_{1}$ and $Y_{2}$) bound
to their receptor sites is determined by parameters $b_{3}$ and $b'_{3}$; 
\item the active state with no bound messenger is determined by parameters
$a_{0}$ and $a'_{0}$; 
\item the active state with messenger $Y_{1}$ bound to the receptor site
is determined by parameters $a_{1}$ and $a'_{1}$; 
\item the active state with messenger $Y_{2}$ bound to the receptor site
is determined by parameters $a_{2}$ and $a'_{2}$; 
\item the active state with messengers ($Y_{1}$ and $Y_{2}$) bound
to their receptor sites is determined by parameters $a_{3}$ and $a'_{3}$. 
\end{enumerate}
We assume that the resting state has a nonnull probability for both
macromolecules. Without loss of generality, we can choose $b_{0}=b'_{0}=1$.
Obviously, if a parameter is nonnull, meaning that the macromolecule
can effectively be in the corresponding state, this implies that there
exists a path joining this state to the default state along with all
intermediary states that have nonnull parameters. For instance, if
$a_{3}>0$, we must have either ($b_{3}>0$ and $b_{1}>0$) or ($b_{3}>0$
and $b_{2}>0$) or ($a_{1}>0$ and $b_{1}>0$) or ($a_{2}>0$ and
$b_{2}>0$). Fortunately, there are various sets of parameters that
fulfill these requirements and that allow us to compute the sum, the
product or the quotient. Here is a possible solution: 
\begin{itemize}
\item for the sum, we can choose $b_{1}=b_{2}=a_{1}=a_{2}=a'_{0}=1$, $b_{3}=a_{0}=a_{3}=b'_{3}=a'_{1}=a'_{2}=a'_{3}=0$,
$b'_{1}=b'_{2}=4$ and $a_{P}n/a_{R}m=1/2$, so that $z=\Phi_{P}(y_{1},y_{2})/\Phi_{R}(y_{1},y_{2})=y_{1}+y_{2}$; 
\item For the product (see Figure \ref{fig:product}), we can choose $b_{1}=b_{3}=a_{3}=a'_{0}=1$, $b_{2}=a_{0}=a_{1}=a_{2}=a'_{1}=a'_{2}=a'_{3}=b'_{2}=0$,
$b'_{1}=2,b'_{3}=4$ and $a_{P}n/a_{R}m=1/2$, so that $z=y_{1}.y_{2}$; 
\item For the quotient, we can choose $b_{1}=b_{2}=a_{1}=1$, $b_{3}=a_{0}=a_{2}=a_{3}=b'_{3}=a'_{0}=a'_{1}=a'_{3}=0$,
$b'_{1}=2,b'_{2}=a'_{2}=a_{P}n/a_{R}m=1/2$, so that $z=y_{1}/y_{2}$. 
\end{itemize}
Hence, by recursion, any RFNC (and any set of RFNCs) can be implemented
by a cascade of theoretical biochemical processes receiving as inputs
a set of primary messengers having concentrations equal (or proportional)
to the coefficients of the input vector $x$.

\section{\textmd{Discussion}}

We proposed to look at an aspect of cell signaling and biochemistry
from a new viewpoint, assuming that these processes may be seen as
performing probabilistic inference.

This proposition is founded on mathematical equivalences between descriptive
probabilistic models of the interactions of populations of macromolecules
and messengers on the one hand, and subjective probabilistic models
of the interaction with the environment on the other hand.

The main necessary hypotheses may be summarized as follows.
\begin{enumerate}
\item We consider a small volume (order of magnitude: $1\mu m^{3}$) of
cell cytoplasm in which thermal diffusion ensures homogeneous concentrations
of messengers at the considered time scale (in the range 1 ms to 1
s). 
\item We consider one or several populations of allosteric macromolecules
with a fixed concentration but variable conformations. 
\item We consider one or several messengers with a single conformation but
varying concentrations. 
\item We consider the conformational changes of the macromolecules, which
control the concentrations of the messengers and reciprocally, the
concentrations of the messengers, which control the conformational
changes of the macromolecules. 
\end{enumerate}
Assuming these four hypotheses, we have demonstrated that descriptive
models of biochemical systems might be seen as performing probabilistic
inference of some well-known and interesting subjective models. Although
the validity of these four hypotheses remains to be discussed, the propositions
made in this paper open new perspectives for future directions of study, for instance: can we propose subjective
interpretations of descriptive models of other types of biochemical
interactions? At the same time and space scale? At different time
and space scales? 

\subsection{\textmd{Validity of the hypotheses?}}

A critical issue for designing descriptive model is to specify the
space and time range in which processes are analyzed.

In the spatial domain, molecular assemblies of nanometer size govern
the behavior of organisms that are several meters long. In this paper, we have
restrained our analysis to small compartments of a single cell, typically
a portion of a dendrite or axon. The order of magnitude of the size
of such a compartment is $1\mu m^{3}=10^{-18}m^{3}$.

Biological events range from about $10^{-12}s$ \cite{Knapp16052006, R:2007:A-Mileston:i}
for the fastest observed conformational changes to millions of years for
gene evolution. We have restrained our analysis to time windows of 1 ms
to 1 s, which is long enough to insure homogeneous concentrations of messengers
in the considered compartments.

\bigskip{}

\noindent Our description of biochemical system is still too schematic
for several reasons.
\begin{itemize}
\item The distinction between macromolecules and diffusible messengers is
not sharp. Macromolecules also diffuse within membranes \cite{Triller:2005:Surface-tr:v}
and their spatial distribution, for instance, between cytoplasmic and
nuclear compartments, is also controlled by specific biochemical mechanisms
\cite{stipanovich:cea-00944345}. However, at the space and time scale we are considering,
this mobility of macromolecules may be neglected as a first approximation. 
\item A number of macromolecules directly act on other macromolecules without
intermediate messengers. A clear example in cell signaling
is given by the interplay of kinases and phosphatases. This was
also discarded from our study for the moment. 
\item Detailed models of biochemical processes should also take into account
the particular geometry of the cell. This includes (3D)  phenomena
arising in homogeneous volumes (such as the concentration of diffusible
messengers), (2D) phenomena arising on a membrane area (such as the distribution
of channels), and (1D) phenomena arising mainly along a symmetry axis
(such as the propagation of potential along dendrite or axon branches).
At present, we have only considered homogeneous concentrations of messengers
and fixed distributions of channels, and we have neglected the effect of
membrane potentials. 
\item We have only considered biochemical systems at equilibrium. Clearly,
the temporal evolution of concentrations and macromolecular configurations
must be further developed, in particular to account for slow reaction
and diffusion processes, and for the various roles of biochemical
feedback pathways. These dynamic processes could be related to
time-evolving probabilistic reasoning, for instance hidden
Markov models and Bayesian filters. However, the wide diversity of
time scales encountered in biological systems contrasts with the somewhat
schematic and oversimplified view of time representation in the usual
subjective models. In future work, promising ideas could be the development
of a more complex temporal hierarchy, and a more subtle view of
the respective roles of memory and temporal reasoning in subjective
models.
\end{itemize}

\subsection{\textmd{Subjective interpretation of other biochemical interactions?}}

Obviously, the above detailed hypotheses are too restrictive. Furthermore,
in this work, we have only considered a few possible biochemical interactions
among the huge variety of possible interactions. An important task in the
near future will be to relax these hypotheses and look for subjective
models that could be associated with other kinds of biochemical interactions.
Some instances of such perspectives are discussed in the sequel to
this section.

\subsubsection{\textmd{At the same time and space scale}}

\paragraph{Macromolecules with more allosteric states:}

The allosteric theory \cite{Monod196588}, initially developed to account
for regulatory enzyme kinetics, postulated that proteins undergo fast,
reversible transitions between a discrete number of conformational
states. Transitions occur spontaneously, but some are favored by fixation
of ligand to specific receptor sites. This model has been successfully
applied to a large number of fundamental macromolecular assemblies,
such as hemoglobin, ionic channels, and nuclear or membrane receptors
(see \cite{Changeux03062005} for a review).

The number of conformational states and the variety of controlled
mechanisms for conformational changes can be relatively high. As an
example, DARPP-32, a key macromolecule in the integration of dopamine
and glutamate inputs to the striatal GABAergic neurons, exhibits four
phosphorylation sites, thus 16 conformational states, and each phosphorylation
is controlled by a different chemical messenger pathway \cite{fernandez:inserm-00705903}.
Inactivation of rhodopsin in vertebrate photoreceptors involves up
to 12 phosphorylation sites, which are all controlled by the intracellular
calcium concentration. In \cite{Houillon2010} we described a first application
of our approach to the vertebrate phototransduction biochemical cascade.
Augmenting the number of conformational states opens very exciting
perspectives on the complexity of the computation that a single population
of macromolecules could perform, but the corresponding subjective
models are still to be proposed.

\paragraph{Macromolecules with several receptor sites for the same messenger:}

Macromolecules with several receptor sites for the same messenger
are very frequent in biochemistry. A number of allosteric macromolecules,
including ionic channels, are composed of several subunits, and some of
them are identical. The presence of a receptor site for a specific
messenger on each subunit makes the whole macromolecule controlled
by the second, third, or fourth power of the concentration. As a consequence,
the catalytic activity is a highly nonlinear function of the messenger
concentration, and may exhibit very sharp sensitivity.

In terms of subjective models, this means that similar observations,
converted into likelihood ratios, are performed several times to infer
the posterior distribution. This might be a simple and elegant way
to enrich the computational complexity performed by macromolecules
without requiring high dynamic ranges of messenger concentration.
In the present work, we have assumed a simple proportional relationship
between likelihood ratios and messenger concentrations. The presence
of multiple receptor sites for the same messenger, and more generally
the multimeric structure of some macromolecules can be interpreted
as a nonlinear coding of the likelihood ratio, which could be better
adapted to the biological constraints. For instance, a tenfold increase
in messenger concentration could correspond to a likelihood ratio multiplied
by $10^{4}$ for a tetrameric receptor.

\paragraph{Allosteric changes governed by other events instead of chemical
messengers:}

Allosteric changes may be caused by other types of events besides chemical
messengers such as electrical or mechanical ones. These events are
not considered in the present model and should be studied in future.

Single-channel voltage clamp recordings \cite{Neher:1976:Single-cha:w, Sakmann:1995:Single-Cha:l} have revealed
several important characteristics of channels: (i) currents through
isolated channels, and thus channel conductance, alternate between
discrete values; (ii) transitions between current/conductance values
are random brisk events; (iii) the transition probabilities can be modulated
by pharmacological and biological agents (neurotransmitters, second
messengers), ions (calcium), or membrane potentials. In agreement with
the allosteric theory, the current descriptive model of ion channels
\cite{Colquhoun06031981, Lauger:1995:Conformati:b, Chung:1996:Coupled-Ma:x} is that of a finite state Markov model,
similar to the one we used in this paper. Some transitions depend
on the presence of a specific messenger in the vicinity of receptor
sites, either in the extracellular domain (ionotropic receptor-channels)
or in the intracellular domain, where receptor sites are specific
to second messengers or calcium transporters. In voltage-dependent
channels, transition probabilities are controlled by the membrane
potential.

\paragraph{Local cascades and feedbacks:}

Metabotropic receptors are transmembrane macromolecular assemblies
with a receptor site in the extracellular domain and an activatable
site in the intracellular domain. Chemical binding of the neurotransmitter
with the extracellular receptor site induces allosteric conformational
change of the macromolecule, which activates the intracellular site
(an activatable enzyme site such as protein kinase, protein phosphatase,
or G-protein release). More generally, macromolecules can be activated
by various events like mechanical strength, photoisomerization, pheromones,
or odor detection. The intracellular activity results in the release
of a primary messenger (\textit{e.g.}, the $G_{\alpha}$ subunit in the G-protein-
dependent signaling), which triggers the production or elimination
of a diffusible second messenger (cyclic nucleotides, inositol triphosphate,
\dots), which in turn acts on ionic channels.

In the complex molecular chain from the primary receptor to
the set of ionic channels, several allosteric macromolecules intervene.
Most of them have receptor sites for calcium, or are calcium transporters
like calmodulin, thus allowing feedback regulation. As discussed
above, the role of feedback should be understood in terms of the dynamics
of the systems, including long-term adaptation.

\subsubsection{\textmd{At different time and space scales}}

\paragraph{Different dynamics to store information:}

In the core of our model, we postulate that the existence of well-
separated fast rates ($10^{4}$ to $10^{5}s^{-1}$) and relatively
slow rates ($10$ to $10^{3}s^{-1}$) of biochemical reactions is fundamental.
It allows the Markov processes of molecular collision and configuration
changes generated by thermal agitation to converge to quasistationary
states that approximate probabilistic inference of subjective models.
We have not yet considered the very slow dynamics of some allosteric
changes (like desensitization) or the diffusion of large molecules in
membranes, among many other slow biochemical events. These mechanisms
can clearly be used for accumulating and storing information over long
periods, which is a key computational capacity for adaptation and learning.

\paragraph{The role of membrane potential at cell scale:}

The overall effect of all ionic fluxes can be summed up in the membrane
potential kinetics. Though it is rather unusual to include membrane
potentials in the set of messengers, it seems appropriate because the kinetic
equations of membrane potential are similar to those for chemical
messenger concentrations. Moreover, the membrane potential controls
macromolecule conformational transition in a similar way, although
constrained to a particular, but important, class of macromolecules,
namely the voltage-gated channels \cite{Bezanilla:2000:The-voltag:p}. In turn, ionic channels
change the membrane potential similarly to the way that activatable enzymes
control chemical messenger concentration. Membrane potentials propagate
much faster than messengers diffuse and can transmit the result of a given local computation to the whole
cell. Consequently, including
membrane potentials into our proposed framework would extend the space
scale from a compartment to the whole cell.

\paragraph{Unicellular organisms:}

At this molecular description level, our proposal applies not only
to brain-controlled complex organisms and not only to small neuron networks,
but also to unicellular organisms. Simple organisms like \emph{Paramecium}
or \emph{Euglena gracilis} have limited numbers of sensors, and a greatly
reduced repertoire of actions, but they must nevertheless adapt
their behavior to an even more unpredictable environment. The efficiency
of probabilistic reasoning with an incomplete model of the world applies
equally to these simple organisms. Furthermore, the biochemical mechanisms
that we propose for implementing probabilistic computation are already
effective in controlling the behavior of eukaryotes. Among many other
examples, it has been shown \cite{Iseki:2002:A-blue-lig:h} that the photoavoidance behavior
of the microalgae (\emph{E. gracilis}) is mediated by concentration changes
of cyclic adenosine monophosphate, a second messenger known to be involved
in olfaction and many neuronal signaling pathways of multicellular
animals.

\paragraph{Information transmission in multicellular organisms:}

Excitable cells, particularly neurons, differentiate from other
cells in complex multicellular organisms, from cnidarians to all bilaterians.
These excitable cells provide a distant and discrete mode of messenger
flux control: both chemical or ionic diffusion and passive electrical
propagation become very ineffective at long distances. Note, however,
that slow, but distant, signal propagation through active biochemical
processes without action potentials has been recently discovered \cite{fasano:inserm-00166490}.
On the contrary, active regeneration of action potential thanks to
voltage-gated channels allows the message to be transmitted unchanged
at high speed along axonal branches up to the presynaptic terminals,
where it is converted into chemical signals. Distant interactions
between macromolecular assemblies by action potential propagation
allow multicellular organisms to reach sizes and speeds far greater
than the limit imposed by passive diffusion process. This constitutes
an obvious gain for long-distance communication.
However as opposed to membrane potential and messengers' concentrations which are graded and local signals, the action potentials are all-or-none signals. Therefore long distance communication is achieved at the cost of signal impoverishment.
According
to our view, fast-spike propagation must be completed by local and
graded signal processing involving complex biochemical interactions
that constitute a fundamental mechanism for probability computation.

\section{\textmd{Materials and Methods}}

\subsection{\textmd{Stationary distributions of Markov models are RFNCs of the
transition matrix coefficients: a formal demonstration}}

\label{Demo}

\subsubsection{\textmd{Decomposition for the stationary distribution of a Markov
model.}}

We consider a set $\{X_{i}\}_{1\leq i\leq n}$ of states with the transition
probabilities among those states $T$. The dynamics for the
Markov chain with initial occupancy $P^{0}$ is thus $P^{t}=TP^{t-1}$:
\[
{P^{t}=\begin{pmatrix}T_{11} &  & \cdots &  & T_{1n}\\
\\
\vdots &  & \ddots & T_{ij} & \vdots\\
\\
T_{n1} &  & \cdots &  & T_{nn}
\end{pmatrix}{P^{t-1}}}
\]
where the probability of a transition from $X_{j}$ to $X_{i}$ is
$T_{ij}$. If the Markov process is finite and irreducible, then according to the Perron--Frobenius theorem, it has a unique stationary distribution $\pi$ \cite{Stroock2014}:

\begin{equation}
\pi_{i}=\frac{det((I-T)^{\{i\}})}{\sum\limits _{k=1}^{n}det((I-T)^{\{k\}})},\label{statdist}
\end{equation}
where $M^{\{k\}}=(I-T)^{\{k\}}$ denotes the matrix
obtained after removing the term $m_{ii}$ and its corresponding \ column
and row. For every $X_{j}$, we have $\sum\limits _{i=1}^{n}T_{ij}=1$.
We note that $s_{j}=1-T_{jj}$ is positive as $s_{j}=\sum\limits _{\substack{i=1\\
i\neq j
}
}^{n}T_{ij}$. 
\[
M^{\{k\}}=\begin{pmatrix}s_{i_{1}} & \cdots & -T_{i_{1}j_{n-1}}\\
\vdots & \ddots & \vdots\\
-T_{i_{n-1}j_{1}} & \cdots & s_{i_{n-1}}
\end{pmatrix}
\]
By the Leibniz formula: $detM^{\{k\}}=\sum\limits _{\sigma\in\mathcal{S}}\epsilon(\sigma)\prod\limits _{\substack{j=1\\
j\neq k
}
}^{n}M_{\sigma(j)j}$, where $\mathcal{S}$ are the permutations over indexes $\{1,\cdots,n\}\backslash{k}$.
Because of the form of the diagonal coefficients, this can be rewritten as:
$detM^{\{k\}}=\sum\limits _{a\in\mathcal{A}^{\{k\}}}\lambda(a)\prod\limits _{\substack{j=1\\
j\neq k
}
}^{n}T_{a(j)j},$ where $\mathcal{A}^{k}$ is the set of applications: $\begin{array}{ccccc}
a & : & \{1,\cdots n\}\backslash k & \to & \{1,\cdots n\}\\
 &  & j & \mapsto & a(j)\neq j
\end{array}$, that is, the set of applications on indexes from $1$ to $n$ except
$k$, leaving no indexes invariant and with $k$ as a possible image.
We consider elementary matrices $^{a}M^{\{k\}}$ with

$\begin{array}{ccccc}
\forall j\in\{1,\cdots n\}\backslash k\\
\forall i\in\{1,\cdots n\}\backslash k & s.t. & i\neq a(j) & ^{a}M_{ij}^{\{k\}}=0.
\end{array}$

The determinant of such a matrix is reduced to: $det^{a}M^{\{k\}}=\lambda(a)\prod\limits _{\substack{j=1\\
j\neq k
}
}^{n}T_{a(j)j}$. Note that the coefficient $\lambda(a)$ is the same for $det^{a}M^{\{k\}}$
and $detM^{\{k\}}$, so that: $detM^{\{k\}}=\sum\limits _{a\in\mathcal{A}^{\{k\}}}det^{a}M^{\{k\}}.$

\subsubsection{\textmd{Almost-diagonal matrices}}

For a matrix (hereafter referred as an almost-diagonal matrix) having positive
coefficients on the diagonal and at most one other single nonzero coefficient
in each column, this additional coefficient being opposite to the
diagonal term, we denote by $K(M)$ the number of columns having
a single nonzero coefficient. The elementary matrices defined above
are almost-diagonal with $K(^{a}M^{\{k\}})=\#\{j:a(j)=k\}$ and applications
$^{a}M^{\{k\}}$ can then be studied based on their $K$
values.

\paragraph{\textmd{If $K=0$}}

$s_{j}=T_{a(j)j}$ and the matrix is reduced to: 
\[
^{a}M^{\{k\}}=\begin{pmatrix}T_{a(1)1} & \cdots &  & \cdots &  & 0 &  & \cdots & 0\\
0 &  &  &  &  & -T_{a(j)j} &  &  & \vdots\\
\vdots &  & \ddots &  &  & \vdots &  &  & -T_{a(n)n}\\
-T_{a(1)1} &  &  &  &  & \vdots &  &  & 0\\
0 &  &  &  &  & T_{a(j)j} &  &  & \vdots\\
\vdots &  &  &  &  & \vdots &  & \ddots & 0\\
0 & \cdots &  &  &  & 0 &  & \cdots & T_{a(n)n}
\end{pmatrix}
\]

For such a matrix, the terms in each column sum to zero  so that
$det^{a}M^{\{k\}}=0$ and then $\lambda(a)=0$.

\paragraph{\textmd{If $K=n-1$}}

In this case, the matrix is diagonal and $det^{a}M^{\{k\}}=\prod\limits _{\substack{j=1\\
j\neq k
}
}^{n}T_{kj}$ so that $\lambda(a)=1$.

\paragraph{\textmd{If $K\in\{1,\cdots,n-2\}$}}

There is at least one column with only its diagonal term nonzero.
After Cramer expansion along one such column (for example the smallest
$j$ such that $a(j)=k$), $det^{a}M^{\{k\}}=T_{kj}det^{a}M^{\{k\}\{j\}}$.
The new matrix $^{a}M^{\{k\}\{j\}}$ is of dimension $n-2$
and is almost diagonal. We name $K_{1}(^{a}M^{\{k\}})\in{0,\cdots,n-2}$
the number of columns having a single nonzero coefficient (on the
diagonal). If $\exists i|a(i)=j$, $K^{1}(^{a}M^{\{k\}})=K(^{a}M^{\{k\}})$,
otherwise $K_{1}(^{a}M^{\{k\}})<K(^{a}M^{\{k\}})$,
we then iterate this operation generating the sequence $(K_{s}(^{a}M^{\{k\}}))_{1\leq s\leq n-2}$.
As the $K_{s}$ are bounded by the matrix dimension $\forall s\in\{1,\cdots,n-2\},K_{s}(^{a}M^{\{k\}})\leq n-s-1$,
there is either an $s=s_{+}$ for which $K_{s^{+}}(a)=n-s-1$ and then
$\lambda(a)=1$ (as shown in the case $K=n-1$) or an $s=s^{-}$ for which
$K_{s^{-}}(a)=0$ and then $\lambda(a)=0$ (as shown in the case $K=0$).

\subsubsection{\textmd{Stationary distributions of Markov models are RFNCs of the
transition matrix coefficients.}}

We showed that the determinants of $M^{\{k\}}=(I-T)^{\{k\}}$
can be decomposed as $detM^{\{k\}}=\sum\limits _{a\in\mathcal{A}^{\{k\}}}det^{a}M^{\{k\}}$
and $\forall a\in\mathcal{A}^{\{k\}},\lambda(a)=1$ or $\lambda(a)=0$,
so these determinants are polynomials in the transition probabilities
with nonnegative coefficients. It then follows from \eqref{statdist}
that the stationary distributions for a Markov chain are RFNCs of the
transition probabilities.

\subsection{\textmd{Simulations}}

\subsubsection{\textmd{Algorithms}}

We performed simulations with a software package written
in C++. To simulate the descriptive models, we used the direct version
of the exact stochastic Gillespie algorithm \cite{Gillespie:1977:Exact-Stoc:r, Gillespie:2007:Stochastic:y}. The
current state of the biochemical system is a vector of natural integers
$X^{t}$. Each component $X_{i}^{t}$ corresponds to the current number
of molecules in a particular conformation in a particular compartment.
The reactions are considered as random events $R^{t}$. Each reaction/event
can occur with an instantaneous rate (or propensity) proportional
to the number of reactants, \textit{e.g.}, if $i$ and $j$ are the index of
the reactants involved in reaction $r$:

\begin{equation}
a(r,t)=c(r)X_{i}^{t}X_{j}^{t}
\end{equation}

The next reaction is drawn randomly from the histogram of reaction
rates:

\begin{equation}
r^{t}\thicksim P(R^{t}=r)=\frac{a(r,t)}{a_{0}(t)}\,,\, with\, a_{0}(t)=\sum_{r}a(r,t)
\end{equation}

The elapsed time at which the next reaction occurs after time $t$
is drawn from the exponential distribution with intensity equal to
the sum of reaction rates:

\begin{equation}
\Delta t\sim a_{0}(t)^{-1}e^{-a_{0}(t)\Delta t}
\end{equation}

The state vector is then updated by the stoichiometric vector of the
reaction:

\begin{equation}
X^{t+\Delta t}=X^{t}+S(r^{t})
\end{equation}

The components of the stoichiometric vector are integers defining
the change in reactant number resulting from the occurrence of a given
reaction. For instance, if the reaction $r$ involves the collision
of reactants $i$ and $j$, which results in the formation of a new
reactant $k$, then we have $S_{i}(r)=S_{j}(r)=-1$ and $S_{k}(r)=+1$.
In addition, some components of the state vector are set to prespecified
values. They constitute the inputs to the biochemical process.

\bigskip{}

\noindent Bayesian models including Bayesian filters are simulated
by computing the exact inference, \emph{i.e.}, by applying the Bayesian rule
and marginalization rules on histogram distributions.

\subsubsection{\textmd{Parameters}}
\label{sec:parameters}

\paragraph{Figure \ref{fig:one-receptor-site}:}

\textcolor{black} {We simulate a population of $N=100$ receptor sites
($R)$ specific for the messenger acetylcholine ($X$) and the following
two reactions:}

\textcolor{black}{Fixation: $R+X\rightarrow RX$ Rate coefficient: $\beta$}

\textcolor{black}{Removal: $RX\rightarrow R+X$ Rate coefficient: $\alpha$}

The integer numbers of molecules of each species
are computed for a diffusion volume of $1\,\mu m^{3}$. The acetylcholine
concentration is fixed to $2\,\mu M$ (\textit{i.e.}, $n\simeq1200$ molecules)
for the first $10\, ms$ then jumps to $20\,\mu M$ for the next $5\, ms$,
then drops back to the initial value. 

For multiple realizations of the simulation, a differential equation for the mean fraction of activated macromolecules can be derived from the master equation (see \cite{Gardner:2004:Handbook-o:k}):

\begin{equation}
\frac{dq(t)}{dt}=-\alpha q(t)+\beta x(1-q(t)).  
\end{equation}

Similarly, the dynamics of the variance is:
\begin{equation}
\begin{split}
\frac{dv(t)}{dt}=\alpha (q(t)-2v(t))+\beta x(1-q(t)-2v(t))\\
\end{split}
\end{equation}

\begin{table}[h]
\renewcommand{\arraystretch}{1.3}
\caption{Parameter values for Figure 1 (Ligand binding.)}
\label{param_table}
\centering
\begin{tabular}{c || c}
\bfseries Parameter name & \bfseries Value\\
\hline\hline
$\beta$: fixation rate   &  $\beta:150\,\mu M^{-1}s^{-1}$\\
$\alpha$: release rate &  $\alpha:8000 s^{-1}$\\
$x_{down}, x_{up}$: ligand concentrations & $x_{down}$: $2 \mu M$, $x_{up}$: $20 \mu M$\\
$m$: size of the population & $m: 100$\\
$N$: Number of realizations & $N:500$\\
$dt$: time step for ODE integration&$dt:0.005 ms$
\end{tabular}
\end{table}

\paragraph{Figure \ref{fig:two-receptor-site}:}

We simulate a population of $m=100$ NACh channel-receptors
with two receptor sites and one catalytic site $Q_0$ and $Q_1$ ($[Q=0]$ and $[Q=1]$). There
are nine molecular species (the ACh messenger plus the eight conformational
states of the channel-receptor), and $24$ reactions. Both receptor
sites have the same fixation ($\beta$) and removal rates ($\alpha_{0}$ and $\alpha_{1}$).  The transition rates between $Q_0$ and $Q_1$ states depend on the number $i$ (0, 1 or 2)
of fixed ligands.


\begin{table}[h]
\renewcommand{\arraystretch}{1.3}
\caption{Parameter values for Figure \ref{fig:two-receptor-site} (acetylcholine receptor.)}
\label{param_table_2}
\centering
\begin{tabular}{c || c}
\bfseries Parameter name & \bfseries Value\\
\hline\hline
$\beta$: fixation rate   &  $\beta:150 \mu M^{-1}.s^{-1}$\\
$\alpha_0, \alpha_1$: removal rates &  $\alpha_0:8000.s^{-1}, \alpha_1:8.64\,s^{-1}$\\
$k^a_m$: activation rate when $m$ ligands are bound & $k^a_0$: $0.54$, $k^a_1$: $130$, $k^a_2$: $30000 s^{-1}$\\
$k^d_m$: deactivation rate when $m$ ligands are bound & $k^d_0$: $10800$, $k^d_1$: $2808$, $k^d_2$: $700 s^{-1}$\\
$x_{down}, x_{up}$: ligand concentrations & $x_{down}: 2 \mu M$, $x_{up}: 20 \mu M$\\
$m$: size of the population& $m:100$\\
$N$: Number of realizations & $N:500$\\
$dt$: time step for ODE integration & $dt:0.005 ms$
\end{tabular}
\end{table}

\paragraph{Figure \ref{fig:product}:}

We simulate the transitions of 200 macromolecules using the Gillespie algorithm. Half of the molecules are involved in the production of $Y$ and may be activated when both receptor sites are occupied and receptor binding is sequential (first $X_1$, then $X_2$ ). The other half are involved in the degradation of $Y$ and may be activated only when none of the receptor sites is occupied. Messenger fixation also occurs in a sequential manner. Fixation rates are all assumed to have the same value $\beta$ and removal rates have the same value $\alpha_{1,2}^{P,R}$, except for the removal of $X_2$ in the producing macromolecule $\alpha_2^R$. Activation and deactivation rates also have the same values $k_{P,R}^a$ and $k_{P,R}^d$ except for the deactivation rate of the removing macromolecule $k_R^d$. The dynamics for the production and removal of $Y$ are simulated through integration of the following differential equation:

\begin{equation}
\tau \frac{dy(t)}{dt}=-\Phi_R(x_1, x_2)y(t)+\Phi_P(x_1, x_2).  
\end{equation}

\begin{table}[h]
\renewcommand{\arraystretch}{1.3}
\caption{Parameter values for Figure \ref{fig:product} (Abstract reaction for product computation.)}
\label{param_table_2}
\centering
\begin{tabular}{c || c}
\bfseries Parameter name & \bfseries Value\\
\hline\hline
$\beta$: fixation rate for \\
ligands& $\beta=1000 \mu M^{-1}.s^{-1}$\\
$\alpha_i^R$: release rates for \\
ligand $x_i$ ($i \in {1,2}$) on the\\
 macromolecule involved in production of $Y$& $\beta_1^P:1000 \mu s^{-1}$, $\beta_2^P:2000 \mu s^{-1}$\\
$\alpha^R$: removal rate for \\
ligands & $\alpha^R:1000 \mu s^{-1}$\\
$k^a_P$: activation rate for the \\
macromolecule involved in production of $Y$  & $k^a_P: 2000 s^{1}$\\
$k^a_R$: activation rate for the \\
macromolecule involved in degradation of $Y$  & $k^a_R: 2000 s^{1}$\\
$k^d$: deactivation rate  & $k^d:1000 s^{-1}$\\
$x_{1}, x_{2}$: ligand concentrations & $x_i \in [1 \mu M, 10 \mu M]$\\
$m$: size of the population& $m: 100$\\
$N$: Number of realizations & $N: 500$\\
$dt$: time step for ODE integration & $dt:0.005 ms$\\
$\tau$: time constant for the dynamics of $Y$ & 10 s 
\end{tabular}
\end{table}


\section{\textmd{Acknowledgments}}

This work is part of the European FP7 project
BAMBI (Bottom-up Approaches to Machines dedicated to Bayesian Inferences).
The authors warmly thank Emmanuel Mazer, Julie Grollier, and Damien
Qurelioz for their fruitful comments.

\bibliography{biochemical-computation}

\section{\textmd{Figure Legends}}


\end{document}